\documentclass[10pt,twocolumn,showpacs]{revtex4}
\usepackage{graphicx}

\def\Box{\kern1pt\vbox{\hrule height 1.2pt\hbox{\vrule width 1.2pt\hskip 3pt 
\vbox{\vskip 6pt}\hskip 3pt\vrule width 0.6pt}\hrule height 0.6pt}\kern1pt}

\def\be{\begin{equation}}
\def\ee{\end{equation}}
\def\bea{\begin{eqnarray}}
\def\eea{\end{eqnarray}}

\begin{document}

%\preprint{astro-ph/yymmddd}

\title{ On the stability of phantom K-essence theories}

\author{L. Raul Abramo}
\email{abramo@fma.if.usp.br}
\affiliation{Instituto de F\'{\i}sica,
Universidade de S\~ao Paulo \\
CP 66318, 05315-970, S\~ao Paulo, Brazil}

\author{Nelson Pinto-Neto}
\email{nelson@cbpf.br}
\affiliation{ICRA -- Centro Brasileiro de Pesquisas F\'{\i}sicas\\
R. Xavier Sigaud 150, 22290-180, Rio de Janeiro, Brazil}

\begin{abstract}
We show that phantom dark energy, if it is described by
a K-essence theory, has three fundamental problems: first, its
hamiltonian is unbounded from below. Second, classical stability
precludes the equation of state from crossing the
``Lambda-barrier'', $w_\Lambda=-1$. Finally, both the equation
of state and the sound speed are unbounded --- the first, from below,
the second, from above --- if the kinetic term is not bounded
by dynamics.
\end{abstract}


\maketitle

\section{Introduction}

Observations of Type 1A Supernovae \cite{SN}, of the cosmic microwave
background radiation \cite{WMAP} and of large-scale structure \cite{SDSS} 
indicate that the universe is
currently experiencing a stage of accelerated expansion.
When the SN data is considered independently of any other osbervations,
the resulting constraints on dark energy parameters
allow (indeed, prefer) values for the equation of state
$w=p_{de}/\rho_{de} \lesssim -1$, where $p_{de}$ and $\rho_{de}$ are
the cosmological pressure and energy density of the dark energy
component \cite{jailson}. However, when cosmic microwave background and
large-scale-structure data are considered as well,
the constraints become much
tighter, and there is not much space to wiggle around $w\simeq -1$.
Nevertheless, much thought has been devoted to the possibility
that superacceleration ($w<-1$) may rule our universe in the near
future \cite{phantoms} and even cause a future spacelike
singularity (``big-rip'') \cite{BigRip}.

In this paper we show that, in the realm of General Relativity,
a non-interacting phantom matter field is a very tough sell indeed.
If we try to build phantom matter out of canonical scalar
fields, we are led to consider a negative
kinetic term in the Lagrangian \cite{phantoms},
which makes the theory unstable ---
classically {\it and} quantum-mechanically \cite{negk,Cline}.
On the other hand, if we enlarge the class of scalar field Lagrangians 
to include those of the type of K-essence \cite{kessence}, then, as we
will presently show,
phantom K-matter \cite{0403157} 
suffers from at least three basic problems.

First, phantom K-matter is quantum-mechanically unstable.
This means that there is always a region in phase space where hard (UV)
excitations of the field possess negative energies. Since there is
nothing that can prevent positive-energy modes from decaying into
negative-energy modes, all matter fields in the universe would decay
instantly through tunneling into the negative-energy modes of such
a scalar field. The theory is sick, and there is no cure.

Second, the equation of state is unbounded from below
($w\rightarrow - \infty$) if the kinetic energy is not bounded.
Moreover, as the equation of state diverges, so
does the sound speed, $c_s^2 \rightarrow \infty$. This means that,
besides potential problems with causality, the frequency of field
oscillations diverges near the big-rip singularity. The only
way to stop running into these divergences is if relativistic dynamics
intervenes to stop the field from running to $w=-\infty$. We will
show in Sec. IV that this is indeed possible.

Third, we give a simple proof of a known result
\cite{LambdaBarrier} which shows
that phantom K-matter, with $w<-1$, cannot cross the ``phantom barrier''
at $w=-1$ if one assumes classical stability, $c_s^2 \geq 0$.
This means that if the equation of state is found to cross this barrier,
then dark energy must be described by other fluid-like models
\cite{IntDE} or through non-minimal couplings to gravity \cite{NMC}.

\section{Phantom dark energy with K-essence matter}

A minimal generalization of the usual canonical Lagrangian is the so-called
$K$-Lagrangian, introduced initially to enlarge the class of inflationary
models \cite{kinflation,Garriga-Mukhanov}. Later, they were also studied
as dark energy models (see, e.g., \cite{kessence,kessence_pheno}.)

Any sensible phantom k-essence matter that can drive the
super-acceleration of the Universe should satisfy three criteria:

\begin{quote}
a) It should be a phantom field: $w\equiv p/\rho <-1$ in some spacetime volume.

b) Classical solutions should be stable: the sound speed $c_s^2$ of
classical perturbations of the field around homogeneous FLRW solutions
cannot be negative.

c) Quantum stability: the Hamiltonian must be limited from
below. Evidently, any system with a Hamiltonian which is unbounded from below
would be instantly destroyed by quantum tunelling of positive-energy
particles into the negative-energy particles. For theories with non-canonical
kinetic terms, the quantum stability is a nontrivial issue.
\end{quote}

In what follows we will assume that the matter sector is represented by
a K-essence scalar field Lagrangian, which has the form:

\be
\label{Lag}
{\cal{L}} = \sqrt{-g} \; F(X)V(\phi) \; ,
\ee
where:

\be
\label{def:X}
X \equiv \frac12 \; \partial^\mu \phi \; \partial_\mu \phi \; ,
\ee
and we use the timelike metric signature, $\eta_{\mu\nu} = {\rm diag}(+,-,-,-)$.
For a canonical scalar field theory, $L=X-V(\phi)$.

The energy-momentum tensor of this fluid reads:
\be
\label{temini}
T_{\mu\nu}=\frac{2}{\sqrt{-g}}\frac{\partial\cal{L}}{\partial g^{\mu\nu}}
=VF'\partial_{\mu}\phi\partial_{\nu}\phi-g_{\mu\nu}VF \; ,
\ee
where a prime denotes a derivative with respect to $X$.
In gaussian coordinates, which can be constructed in finite regions
in general and globally if the topology of spacetime is
$M^4=R\bigotimes M^3$, one can write:
\be
\label{X}
X \equiv \frac12 \; g^{00}\partial_0 \phi \; \partial_0 \phi -
\frac12 |g^{ij}| \partial_i \phi \; \partial_j \phi\equiv X_t-X_s\; ,
\ee
where $X_t$ and $X_s$ are both non-negative. The Hamiltonian then reads:
\be
\label{ham}
{\cal{H}}=T^0_0=V\, \left(
2 F'[\Pi,X_s,\phi] X_t[\Pi,X_s,\phi] - F[\Pi,X_s,\phi]
\right)
\; ,
\ee
where $\Pi\equiv F' V \partial^0 \phi$ is the conjugated momentum.

In homogeneous and isotropic spacetimes $X_s=0$, and we obtain:
\be
\label{rho}
\rho=T^0_0=V(2F'X-F)\; ,
\ee
and:
\be
\label{p}
p=-\frac{1}{3} T^i_i=VF\; ,
\ee
implying that:
\be
\label{w}
w=\frac{p}{\rho}=-\frac{1}{F^2\biggl(\frac{X}{F^2}\biggr)'}.
\ee
Hence, condition (a), $w<-1$, yields:
\be
\label{cond}
0< 2 X \frac{F'}{F} < 1 \; ,
\ee
or, equivalently:
\bea
\nonumber
F'>0 \quad , \quad F-2XF'>0 \quad &{\rm if}& \quad F>0 \; , \\
\label{cond2}
F'<0 \quad , \quad F-2XF'<0 \quad &{\rm if}& \quad F<0 \; .
\eea

The sound speed, $c_s^2$, is the function appearing before spatial
gradients in the scalar field equation-of-motion,
$\ddot\phi+c_s^2\nabla^2\phi + \ldots = 0$, and for perturbations
around homogeneous solutions, it is a function of time.
The sound speed expresses the phase velocity of the
inhomogeneous perturbations of the scalar field.
Therefore, to avoid exponentially growing solutions and thus
ensure classical stability, we must have $c_s^2 \geq 0$.
On the other hand, to
ensure causality in the usual sense the condition would be
$c_s^2 \leq 1$. However, since the theories we consider
are perfectly Lorentz invariant (the superluminal propagation being just
a consequence of the nonlinearity of the theory \cite{Garriga-Mukhanov}),
we will only
impose the first condition (classical
stability), as superluminal propagation cannot be ruled out by
observations if the scalar field is dark, i.e., if it does not
interact with normal matter.

For K-essence models the sound speed 
takes the simple expression \cite{Garriga-Mukhanov}:
\be
\label{cs}
c_s^2=\frac{p'}{\rho'}=\frac{{F'}^2}{({F'}^2 X)'} \; .
\ee
Condition (b), that $c_s^2\geq 0$, implies:
\be
\label{cond3}
({F'}^2 X)'\geq 0\; .
\ee
In the subsequent subsections we will
use the following relation between $c_s^2$ and $w$
coming from Eqs. (\ref{w}) and (\ref{cs}):
\be
\label{rel}
1-\frac{w}{c_s^2}=\frac{2Xw'}{1+w}\; .
\ee

Finally, condition (c) implies that ${\cal{H}}=V(2F'X_t-F)$
must be bounded from below. As we will show in the next
section, this is impossible given the two conditions above,
Eqs. (\ref{cond2}) and (\ref{cond3}).

\section{General results}

\subsection{The general Hamiltonian is not bounded from below}

Suppose initially that $V>0$, so the function ${\cal{H}}/V$ must
be shown to be bounded from below. The conditions in Eqs. (10) are 
constraints
on the function $F(X)$ and its derivative, that do not depend
on how $X$ is obtained from the phantom field (or whether it is
 homogeneous or not). That condition must be satisfied whenever
 $X>0$ (which is the only constraint the restriction to homogeneous 
 fields imposes on the possible values of the real variable $X$),
 otherwise the homogeneous model would not yield a super-accelerated 
 expansion, and it should be discarded. As $\rho+3p=2V(F'X+F)<0$ , 
in order to have acceleration in the homogeneous 
 background (where $X>0$), one must take from condition (10) the 
 alternative $F<0$ and $F'<0$ because we are now considering $V>0$.
 Hence, in this case, one must have from Eq. (10) that 
 $F_0'=F'(X_0)<0$ and $F_0=F(X_0)<0$ for any given value $X=X_0>0$.
 We will show below that the subset of fluctuations that keeps $X>0$
constant, while varying $X_t$ and $X_s$, is such that 
the Hamiltonian
 is unbounded from below. This suffices to prove that the system is
 unstable.

Consider, then, quantum fluctuations such that $X$ is kept fixed,
$X=X_t-X_s=X_0>0$, but where $X_t$ can vary freely.
In order for the Hamiltonian to
be bounded from below, one must have $2 X_t F_0'-F_0 > C$, where
C is some finite constant. However, since $F_0'<0$ and $F_0<0$ are
fixed, the stronger the quantum fluctuation, the larger $X_t$ and 
therefore the Hamiltonian is unbounded from below.

On the regions where $V<0$ the proof is similar.
Using the same reasoning, one now has that $F_0'>0$ and $F_0>0$,
and hence ${\cal{H}}=V(2F'X_t-F)$ is unbounded from below 
because here $V$ is negative.
Notice that for $VF<0$, in the case of FLRW backgrounds,
the condition (\ref{cond}) implies that $\rho=-VF^3(X/F^2)'>0$.
Notice also that, within this class of Lagrangians,
the Hamiltonians of the canonical and the Born-Infeld
cases, respectively $F=X$ and $F=-\sqrt{1-2X}$,
{\it are} bounded from below. However, these cases do not
represent phantom fields.

\subsection{The equation of state and sound speed are unbounded
if $w<-1$ and $c_s^2\geq 0$}

From Eq. (\ref{rel}) one has:

\be
\label{rel3}
c_s^2=\frac{|w||1+w|}{2X|w'|-|1+w|}\; .
\ee
In order for $0 \leq c_s^2 < \infty$ we must have:

\be
\label{lim1}
\lim_{X\rightarrow \infty}[-2Xw'(X)]>\lim_{X\rightarrow \infty}|1+w|>0\; .
\ee

We will prove by contradiction that one cannot have
$w(X)=w_{\infty}+f(X)$, where $-\infty < w_{\infty}<-1$ is a constant
and $\lim_{X\rightarrow \infty} f(X) = 0$. Notice that:
\bea
\nonumber
\lim_{X\rightarrow \infty} [-2Xf(X)]' &=&
2 \lim_{X\rightarrow \infty} [-f(X)-Xf'(X)] \\
\nonumber
&=& \lim_{X\rightarrow \infty} [-2Xf'(X)]\\
\nonumber
&>&|1+w_{\infty}| \; ,
\eea
where we used Eq.(\ref{lim1}) and
$\lim_{X\rightarrow \infty} w(X) = w_{\infty}$
in the last step. Hence:
\bea
\nonumber
\lim_{X\rightarrow \infty}f(X) &=&
- \frac12 \lim_{X\rightarrow \infty}[\frac{1}{X} \int (-2X f'(X))dX] \\
\nonumber
&<& - \frac12 |1+w_{\infty}|<0 \; ,
\eea
contradicting the hypothesis that
$\lim_{X\rightarrow \infty} f(X) = 0$.
Hence, for $w<-1$ and $0 \leq c_s^2 < \infty$ we must have that:
\be
\label{wlim}
\lim_{X\rightarrow \infty} w(X) \rightarrow -\infty  \; .
\ee

Similarly, we can prove that the sound speed, $c_s^2$, is also
unbounded (from above) in phantom K-matter models.
Here it is useful to write Eq. (\ref{rel}) as:

\be
\label{rel4}
c_s^{-2}= \frac{1}{w}-\frac{2Xw'}{1+w} \; .
\ee
One can see that $c_s^2$ does not diverge if and only if
the expression above does not goes to zero when
$X\rightarrow \infty$, where, as we have proven
above, $w\rightarrow -\infty$. Hence, for very large values of $X$
we have:
\be
\label{den}
\frac{1}{w}-\frac{2Xw'}{1+w} \approx
2\biggl(\frac{X}{w}\biggr)'
= h(X) = h_{min} > 0\; ,
\ee
where $h_{min}$ is the minimum value of $h(X)$ for $X$ very large.
However, from Eq.(\ref{den}), one could then write:
\be
\label{dem}
\lim_{X\rightarrow \infty}\frac{2}{w}=\frac{1}{X}\int h(X) dX > h_{min} \; ,
\ee
which is contradiction with the previous result that
$\lim_{X\rightarrow \infty} w(X) \rightarrow -\infty$.
Hence, $\lim_{X\rightarrow \infty} c_s^2(X) \rightarrow \infty$.

\subsection{The equation of state cannot cross the value $-1$ if
$c_s^2\geq 0$}

Let us first consider cosmological solutions, for which $X=X_t>0$.
If $w<-1$, Eq.(\ref{rel}) reads:
\be
\label{rel1}
1+\frac{|w|}{c_s^2}=-\frac{2Xw'}{|1+w|}\; ,
\ee
and $w'<0$. This means that, as a function of $X$ the equation
of state is a monotonically decreasing function.

On the other hand, if $-1<w<0$ then Eq.(\ref{rel}) reads:
\be
\label{rel2}
1+\frac{|w|}{c_s^2}=\frac{2Xw'}{|1+w|}\; ,
\ee
and $w'>0$. In this case, $w$ as a function of $X$ is a monotonically
increasing function. Hence, $w$ cannot cross the value $-1$
if the sound speed is non-negative \cite{LambdaBarrier}.

\section{Toy model}

In this section we present an example of a K-essence Lagrangian
where conditions (a) and (b) of Sec. II are satisfied, and in which
no divergences of $w$ and $c_s^2$ appear. The kinetic function
$F(X)$ reads:
\be
\label{toy}
F(X)=1+aX(1+X)-b(1-X^{n})^{n}\; ,
\ee
where $a$ and $b$ are some constants, $n$ is some positive half-integer,
and $V(\phi)$ is left arbitrary.
General-relativistic dynamics
constrain $X$ to the interval $0<X<1$, because at $X=0$ and $X=1$
there are singularities in the derivatives of the curvature tensor.
With $n=5/2$, $a=0.1$ and $b=0.01$, for example, the equation of state remains
bounded, $-1<w<-2$, and the sound speed is always positive but finite,
$0.43<c_s^2<1$ --- see Fig. 1.
However, the parameter subspace where this
occurs seems very small when compared to the full parameter space of this
toy model.

\begin{figure}
\includegraphics[width=8cm]{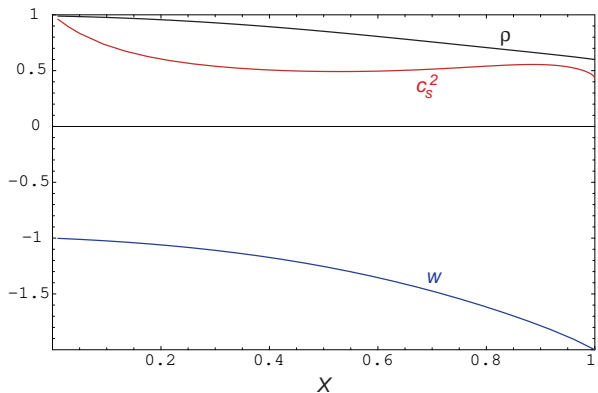}
\caption{\label{fig:1}
The three curves are, from top to bottom: $\rho/V$ (black), the sound speed
$c_s^2$ (red), and the equation of state $w$ (blue).
}
\end{figure}

\section{Discussion}

We have shown in this paper that one cannot obtain K-essence
phantom models without quantum instabilities. This is a direct
consequence of the phantom imposition, namely,
$w=p/\rho\leq -1$ for some range of
$X\equiv \partial^\mu \phi \partial_\mu \phi /2$,
which in fact must be true
for all values of $X$ if one imposes classical stability. Hence,
lagrangians of the type of Eq. (\ref{Lag}) cannot be considered
as fundamental descriptions of phantom fields, being at most
effective lagrangians aplicable to the cosmological set-up.

We have also shown that classically
stable K-essence phantom models present divergences of $w$
in the negative direction ($w$ is bounded from above by $w=-1$,
but unbounded from below), and of $c_s^2$ in the positive direction.
These divergences occur
unless $X\rightarrow\infty$ is dynamically forbidden --
either by limiting the range of $X$ in the lagrangian or by
adjusting the potential $V(\phi)$.
We have exhibited an example of a phantom K-essence lagrangian
with the first property.

\vspace{1.0cm}

{\bf Acknowledgments:} We thank FAPESP, CNPq and CAPES (Brazil)
for financial support. We also thank Giuseppe Dito and
Slava Mukhanov for valuable discussions.

\end{document}